\documentclass[conference]{IEEEtran}
\IEEEoverridecommandlockouts
\pdfoutput=1

\usepackage{cite}

\usepackage{graphicx}
\ifCLASSINFOpdf
\else
\fi
\usepackage[cmex10]{amsmath}
\usepackage{listings}
\usepackage{courier}
\lstnewenvironment{code}[1][]%
{
              \noindent
              \minipage{\linewidth}
              \medskip
              \lstset{frame=lrtb,captionpos=b,basicstyle=\ttfamily\footnotesize,breaklines=true,frame=single,#1}}
{\endminipage}

\usepackage{array}
\usepackage[caption=false,font=footnotesize]{subfig}
\usepackage{multirow}
\usepackage{textcomp}

\makeatletter
\def\footnoterule{\relax%
   \kern-5pt
   \hbox to \columnwidth{\hfill\vrule width 0.5\columnwidth height 0.4pt\hfill}
   \kern4.6pt}
\makeatother

\author{Lauren Milechin}

\begin{document}
%
\title{Database Operations in D4M.jl}

\author{\IEEEauthorblockN{Lauren Milechin\IEEEauthorrefmark{1}, Vijay
    Gadepally\IEEEauthorrefmark{2},   Jeremy Kepner\IEEEauthorrefmark{2}}

\IEEEauthorblockA{\IEEEauthorrefmark{1}MIT EAPS, Cambridge, MA 02139
  \\ lauren.milechin@mit.edu}
\IEEEauthorblockA{\IEEEauthorrefmark{2}MIT Lincoln Laboratory, Lexington, MA 02420}
\thanks{This material is based upon work supported by the National Science Foundation under Grant No. DMS-1312831. Any opinions, findings, and conclusions or recommendations expressed in this material are those of the author(s) and do not necessarily reflect the views of the National Science Foundation.}
}


%


\maketitle

\begin{abstract}
Each step in the data analytics pipeline is important, including database ingest and query. The D4M-Accumulo database connector has allowed analysts to quickly and easily ingest to and query from Apache Accumulo using MATLAB\textregistered{}/GNU Octave syntax. D4M.jl, a Julia implementation of D4M, provides much of the functionality of the original D4M implementation to the Julia community. In this work, we extend D4M.jl to include many of the same database capabilities that the MATLAB\textregistered{}/GNU Octave implementation provides. Here we will describe the D4M.jl database connector, demonstrate how it can be used, and show that it has comparable or better performance to the original implementation in MATLAB\textregistered{}/GNU Octave.
\end{abstract}


%
\IEEEpeerreviewmaketitle

\section{Introduction}
\label{sec:intro}

A database is an integral part of the data analytics pipeline. Datasets that are too large to fit into memory and difficult to organize and search require the use of a database for storage and indexing. For efficient workflows, it is therefore important to not only provide fast ways to get data in and out of the database, but to do so in the language that the analyst is using to analyze their data. This requires both a high-performance database and good connectors that the analyst can use without the burden of learning a new language just to access their data. The D4M (Dynamic Distributed Dimensional Data Model) library, in addition to being a powerful analytic framework, provides database connectors to a number of databases, including Apache Accumulo, a high-performance NoSQL database.

Apache Accumulo is a key-value data store modeled after Google Bigtable \cite{accumulo-webpage}. Some of its features include the ability to add server-side iterators to do custom in-database operations and cell-level visibility labels to restrict access on the individual entry level. Because of its features and performance, Accumulo has been used in a variety of applications, including cloud monitoring \cite{accumulo-cloud}, spatial data  \cite{accumulo-spatial}, and graph processing \cite{newsqlgraphulo2016}. Accumulo has been thoroughly benchmarked, and has shown high performance for both query and ingest \cite{accumulo-bench} \cite{ingest2014}.

The D4M tool is an analytical library for both MATLAB\textregistered{}/GNU Octave and Julia that allows flexible data representation and manipulation \cite{d4m2012} \cite{julia2016}. D4M uses a mathematical structure called an Associative Array to represent data. Associative arrays can represent many different types of data, including graphical, numeric, and string data. They also support a variety of arithmetic and set operations that are facilitated through D4M and have a wide variety of uses \cite{accumuloEigensolver2015} \cite{cmd2015}. These properties make D4M a good fit for large datasets, particularly those that may need to be stored in a database, such as Accumulo.

In order to access these datasets, D4M includes database connectors that allow users to ingest data to and query data from a database. In addition to SQL and SciDB connectors, D4M has a custom database connector for Accumulo that has been available in the MATLAB\textregistered{}/GNU Octave D4M package, providing a simple means to bind to tables for ingesting and querying data. D4M also has a schema that works particularly well for Accumulo \cite{d4mschema2013}. Accumulo was built with fast ingest and query in mind, and past work has shown record-breaking ingest performance using the D4M's ingest and schema \cite{ingest2014}.

The D4M package for Julia developed recently brought many of the capabilities of D4M to the Julia community. Julia is a rapidly growing new language developed for both performance and productivity, providing the ease of use of a high level language, without compromising performance \cite{juliapage}. There are a number of Julia packages that interact with a variety of databases \cite{juliadatabases}, however any work to show performance of these connectors is difficult to find. The initial D4M.jl provides the Associative Array representation and operations, and has been shown to have equivalent or better performance than the original MATLAB\textregistered{}/GNU Octave implementation \cite{julia2016}. In this work, we extend D4M.jl to include database connectors for Accumulo and show how performance compares to the MATLAB\textregistered{}/GNU Octave version for query and ingest.

In the following sections, we will describe these additions. Section \ref{sec:d4m} will introduce D4M more in depth. In Section \ref{sec:juliaimp} we describe the D4M.jl connector. Section \ref{sec:expsetup} will describe the tests we ran to compare ingest and query in the Julia and MATLAB\textregistered{}/GNU Octave D4M implementations, and Section \ref{sec:results} will present and discuss the results of these experiments.

\section{D4M}
\label{sec:d4m}

D4M is open-source software that provides a convenient mathematical representation of the kinds of data that are routinely stored in spreadsheets and large key-value databases. Associations between multidimensional entities (tuples) using string keys and string values can be stored in data structures called associative arrays. For example, in two dimensions, a D4M associative array entry might be:

\vspace{6pt}
\centerline{\textbf{A}(\textquotesingle alice \textquotesingle, \textquotesingle bob \textquotesingle) = \textquotesingle cited \textquotesingle \hspace{4pt} or \hspace{4pt} \textbf{A}(\textquotesingle alice \textquotesingle, \textquotesingle bob \textquotesingle) = 47.0}
\vspace{6pt}

The above tuples have a 1-to-1 correspondence with their key-value store representations:

\vspace{6pt}
\centerline{(\textquotesingle alice \textquotesingle,\textquotesingle bob \textquotesingle,\textquotesingle cited \textquotesingle) \hspace{4pt} or \hspace{4pt}  (\textquotesingle alice \textquotesingle,\textquotesingle bob \textquotesingle,47.0)}
\vspace{6pt}

\begin{figure}[hh]
  \includegraphics[width=20pc]{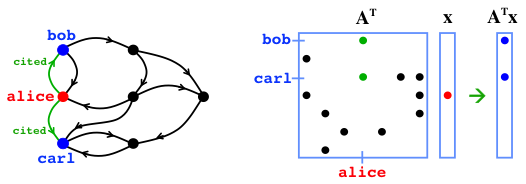}
  \caption{A graph describing the relationship between alice, bob, and carl (left). A sparse associative array A captures the same relationships (right). The fundamental operation of graphs is finding neighbors from a vertex (breadth-first search). The fundamental operation of linear algebra is matrix vector multiply. D4M associative arrays make these two operations identical.  Thus, algorithm developers can simultaneously use both graph theory and linear algebra to exploit complex data.}
  \label{fig:d4mfig}
\end{figure}

Associative arrays can represent complex relationships in either a sparse matrix or a graph structure (see Figure~\ref{fig:d4mfig}). Thus, associative arrays provide a natural data structure for performing both matrix and graph operations. Such algorithms are the foundation of many complex database operations across a wide range of fields~\cite{kepner2011}. Constructing complex composable query operations can be expressed by using simple array indexing of the associative array keys and values, which themselves return associative arrays:

\vspace{6pt}
\begin{tabular}{l l}
\centering
\textbf{A}(\textquotesingle alice \textquotesingle,:) & alice row \\
\textbf{A}(\textquotesingle alice bob \textquotesingle,:) & alice and bob rows \\
\textbf{A}(\textquotesingle al* \textquotesingle,:) & rows beginning with al \\
\textbf{A}(\textquotesingle alice : bob \textquotesingle,:) & rows alice to bob \\
\textbf{A}(1:2, :) & first two rows \\
\textbf{A} == 47.0 & subarray with values 47.0
\end{tabular}
\vspace{6pt}

The composability of associative arrays stems from their ability to define fundamental mathematical operations whose results are also associative arrays. Given two associative arrays A and B, the results of all the following operations will also be associative arrays: 

\vspace{6pt}
\begin{tabular}{c c c c c}
\textbf{A} + \textbf{B} \hspace{5pt}& \textbf{A} - \textbf{B} \hspace{5pt}& \textbf{A} \& \textbf{B} \hspace{5pt}& \textbf{A} $\mid$ \textbf{B} \hspace{5pt}& \textbf{A} * \textbf{B} \\
\end{tabular}
\vspace{6pt}

Measurements using D4M indicate these algorithms can be implemented with a tenfold decrease in coding effort when compared to standard approaches~\cite{d4m2012}.

\section{D4M.jl and Accumulo}
\label{sec:juliaimp}

When D4M.jl was implemented in \cite{julia2016}, the database connectivity feature was left for future work. Here we first give a short description of the initial D4M.jl implementation, followed by the details of the new database features. 

\subsection{D4M.jl}
D4M can be implemented in any language that provides support for sparse linear algebra operations, and this includes Julia. Julia is a newer language developed for both high performance and high-level dynamic programming \cite{juliaFresh2014}. Typically, languages are either high performance, low level, but difficult for development, or high-level and easy for development, but without the performance that a low-level language usually provides. The Julia developers aimed to create a language that is both high-level and high performance. Julia contains both state of the art numeric computation libraries and a state of the art Just-in-Time (JIT) compiler built on Low Level Virtual Machine (LLVM) \cite{juliaLang2012}. In this way, Julia's low-level functionality can be optimized, so that the user does not need to resort to using other languages as low-level building blocks. By leveraging the selected chain of modern programming language technologies within Julia, the Julia community has been rapidly expanding the high-level functions of Julia without compromising in performance. Further, Julia has been shown to be effective in high performance computing \cite{juliaHPC2016}.

D4M.jl provies the functionality of D4M, but with familiar Julia syntax and conventions to the Julia programmer, much the same way D4M-Matlab is designed to be intuitive to the MATLAB\textregistered{} user \cite{d4mjl2018}. This makes D4M.jl more useable to the Julia community. For example, because Julia uses square brackets for indexing rather than the parentheses that are used in MATLAB\textregistered{}, D4M.jl follows this convention.

Benchmarking work has shown D4M.jl to have comparable performance to MATLAB\textregistered{}-D4M, and in some cases surpass the D4M-Matlab implementation in key D4M operations, including matrix multiply and addition \cite{julia2016}. Julia has better support for arrays of strings than MATLAB\textregistered{} does, yielding a simpler code base and better performance on operations Associative Arrays that have string row and column keys.

\subsection{D4M.jl Accumulo Connector}

\begin{figure}[]
{\includegraphics[width=3in]{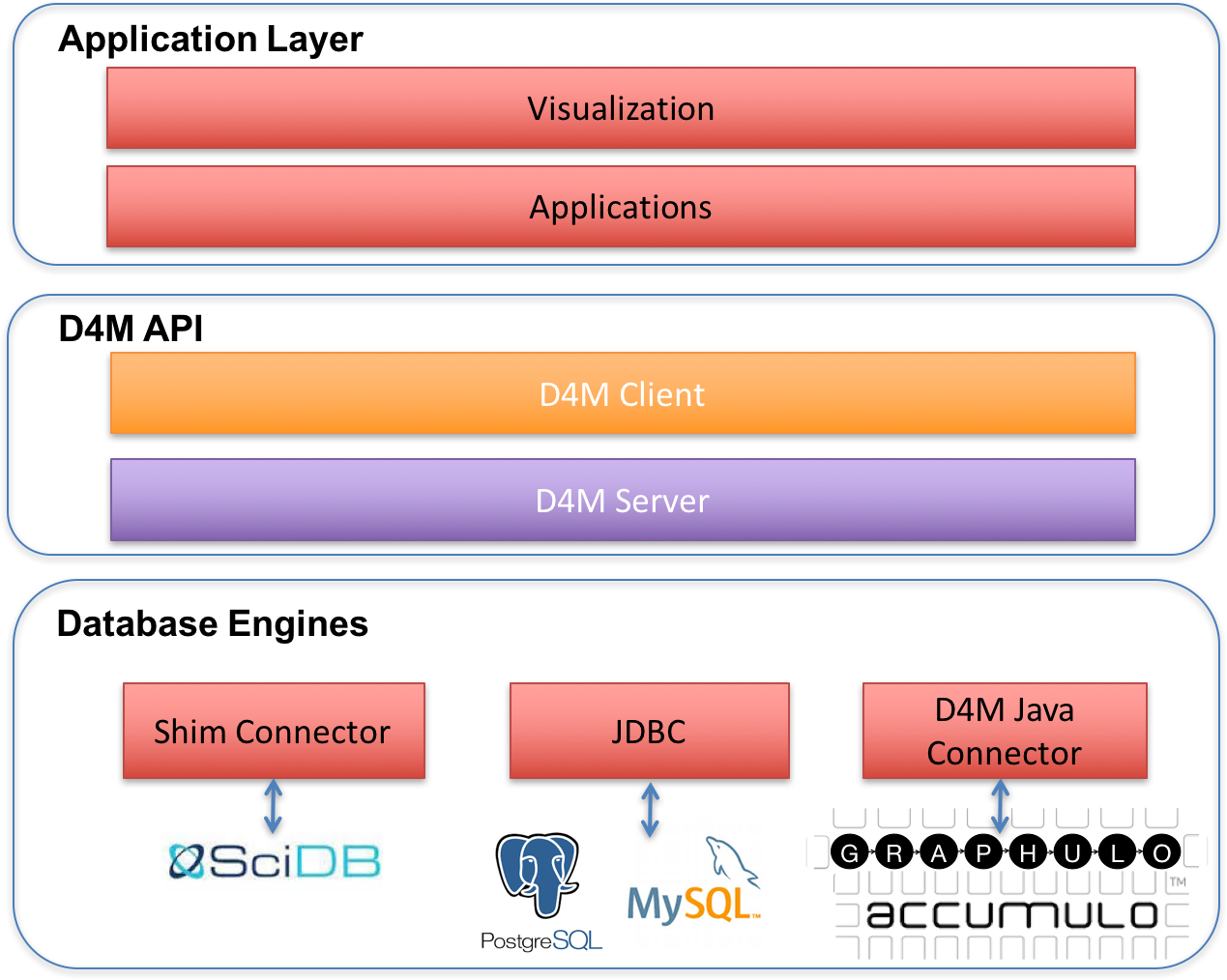}}
\centering
\caption{D4M architecture. D4M server bindings leverage various database connectors, including the custom-built Accumulo connector.}
\label{fig:d4mArch}
\end{figure}

In addition to  the library that provides the Associative Array data structure and its associated operations, D4M gives the user a common syntax to access to several database connectors, including JDBC for SQL and the Shim connector for SciDB. The connector for Accumulo is custom-built in Java and is part of the D4M distribution.

The Java Accumulo connector can be invoked from any language that can call Java functions. In the original D4M implementation, we used Matlab's inherent ability to create Java objects and call Java functions. While Julia natively does not have an ability to call Java functions, the JavaCall.jl package provides this capability \cite{javacall}. JavaCall uses the Java Native Interface to create an in-process Java Virtual Machine (JVM), which is accessed through the jcall function. 

Using JavaCall.jl generally involves importing a class, creating an object, then calling the function you are interested in on that object. At times it can be difficult getting the Java function calls just right, with inputs and outputs of the correct data type, where sometimes Julia data types are sufficient and occasionally Java objects are required. In D4M.jl, we provide an easy-to-use interface to call the Java functions involved in database operations so the user does not have to worry about these details. A number of functions and structs hold java objects and make Java calls. 

\begin{code}[frame=single,label={code:dbcmds}, caption=Using D4M.jl for Accumulo database operations.]
# Initialize JVM
dbinit()

# Connect to Database
DB = dbsetup("mydb02","db.conf")

# Create Tables
Tedge = DB["my_Tedge","my_TedgeT"]
TedgeDeg = DB["my_TedgeDeg"]

# Insert Associative Array into Database
put(Tedge,A)

# Query Database
Arow = Tedge["e1,",:]
Acol = Tedge[:,"v1,"]

# Delete Tables
delete(Tedge)
delete(TedgeDeg)
\end{code} 

The D4M.jl workflow for interacting with Accumulo is as follows. First, a call to \texttt{dbinit()} will initialize the JVM with the required libraries on the class path. The \texttt{dbsetup()} function creates a \texttt{DBserver} struct, which holds the connection information for the Accumulo database, and tables can be created or connected to by indexing into the \texttt{DBserver} struct. Tables can either be single tables or table pairs, which bind to both a table and its transpose, which occur frequently in Accumulo schemas. Data can be ingested using the \texttt{put()} function, which ingests Associative Arrays, or \texttt{putTriple()}, which will ingest arrays of strings. Tables can be queried by using the same indexing syntax as Associative Arrays, and column queries on table pairs will automatically query the transpose table for speed. Finally, tables can be easily deleted using the \texttt{delete()} function. See Listing \ref{code:dbcmds} for an example of this workflow.

\section{Experimental Setup}
\label{sec:expsetup}

While D4M.jl and Matlab-D4M use the same Java Accumulo connector, there is always some small overhead calling these connectors from another language with easy-to-use wrappers. Therefore, we ran some tests to demonstrate the efficiency of D4M.jl compared to Matlab-D4M. We focused on the two most frequently used database operations: ingest and query.

\subsection{Database Ingest}
\label{subsec:ingest}
While data ingest is not the most frequently used database operation, it is the most time consuming. Therefore, it is important to minimize overhead when ingesting data. To compare ingest rates between D4M.jl and Matlab-D4M, we ingested graphs of various sizes with a varying number of ingestors.

Since the best Accumulo ingest rates require having a number of ingest processes inserting data at once, we ran ingest on 1, 2, 4, 8, and 16 processes. For Matlab-D4M, parallel ingest was achieved using the pMatlab library \cite{pmatlab}, and for D4M.jl the SPMD submodule from the DistributedArrays package was used \cite{distarrays}. The SPMD programming model provides the control needed to ensure each process is inserting data at the same time to achieve peak overall ingest rate. Graphs were generated on each of the $k$ ingest processes using the Graph500 unpermuted power law graph generator \cite{d4mpowerlaw2015} with scale ($s$) 12-18 and an average degree ($d$) of 16, producing graphs with $2^s$ vertices and $d*2^s$ edges on each ingest process, or $k*d*2^s$ edges in the final ingested graph. We ingest the adjacency matrices of these graphs and their transposes. Performance is measured in edges ingested/second .

\subsection{Database Query}
The most frequently used database operation is database query. Querying the database is often done interactively, and a slow query response can interrupt an analysts's workflow. To compare query rates between D4M.jl and Matlab-D4M, we queried a large graph for vertices with varying return sizes.

\begin{figure*}[!t]
{\includegraphics[width=3.4in]{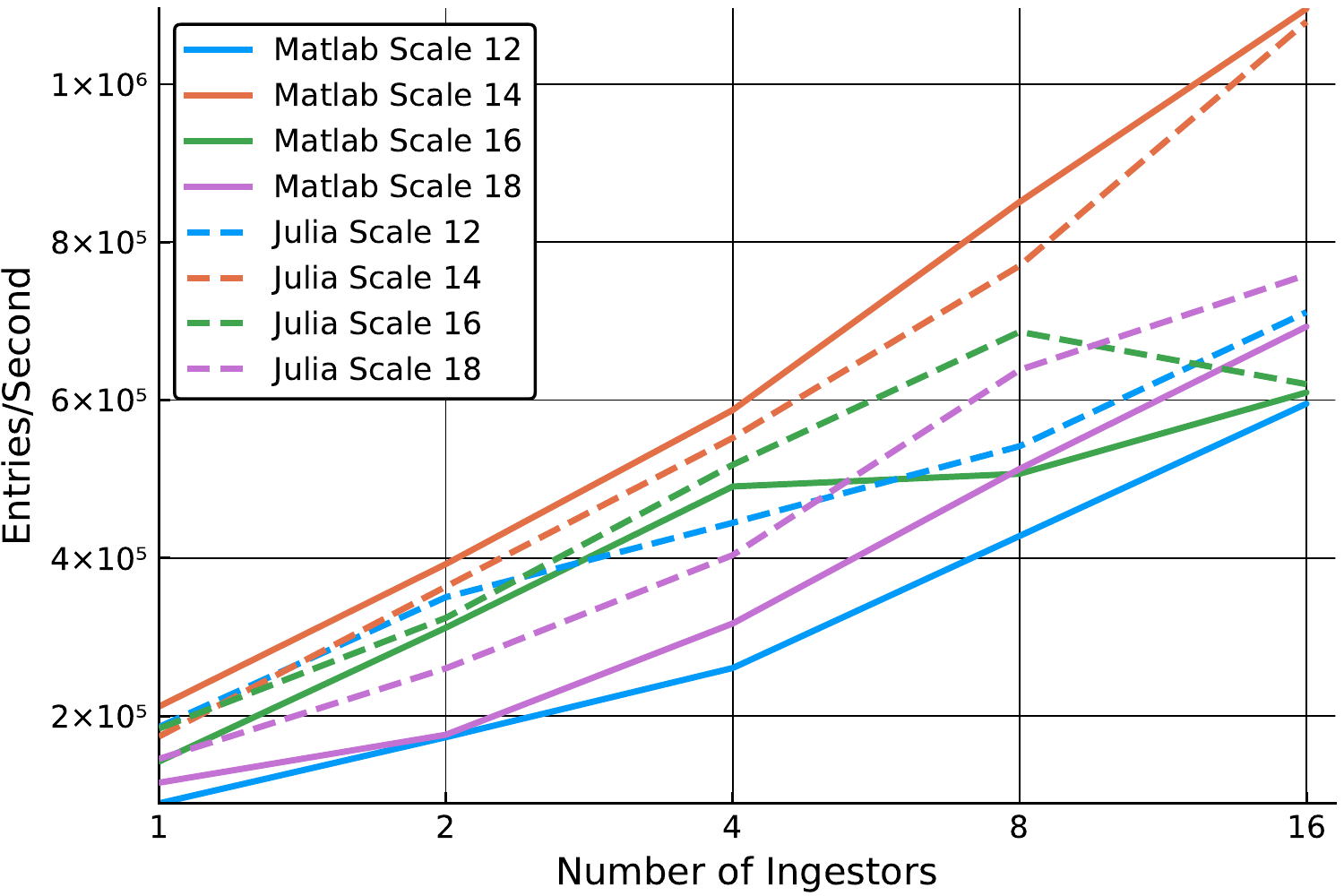}
\label{fig:ingest_cores}}
{\includegraphics[width=3.45in]{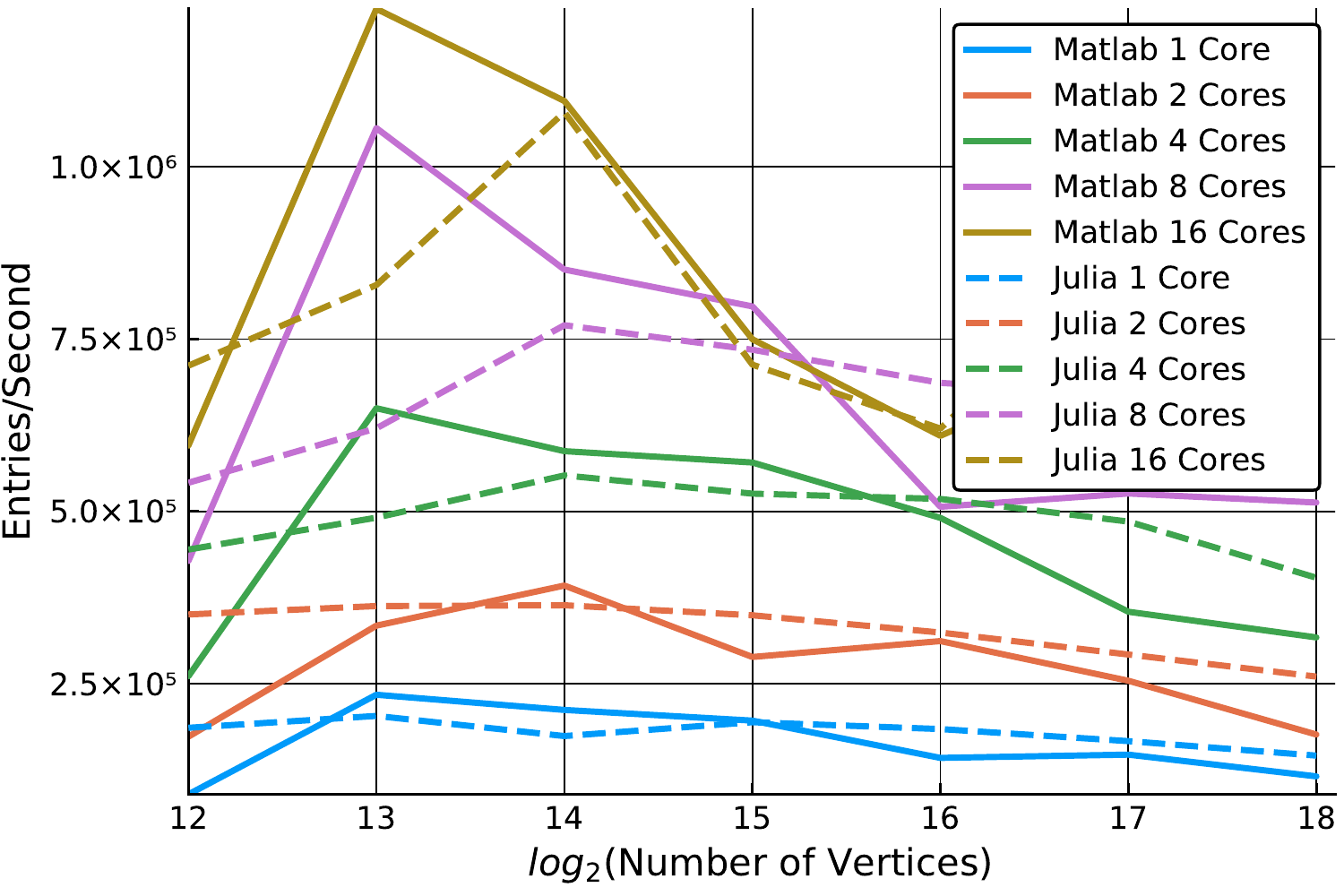}
\label{fig:ingest_scale}} \\
\caption{Ingest rates for Matlab-D4M (solid lines) and D4M.jl (dashed lines). The plot on the left shows ingest rate against the number of ingest processes for several graph sizes, and the plot on the right shows ingest rate against graph size.}
\label{fig:ingestResults}
\end{figure*}

First we ingested a large graph, generated in the manner described in \ref{subsec:ingest}, using 8 processes each ingesting a scale 17 graph (approximately $8*16*2^{17} = 16,777,216$ edges) and its degrees in a separate Degree Table. Using the Degree table, we find which vertices have in and out degrees of approximately size 1, 10, 100, 1000, and 10000. Each query is run on random chosen vertices from these categories, with vertices kept consistent between the Matlab-D4M and D4M.jl queries. Four types of queries were run: single and multiple vertex row and column queries. The multiple vertex queries selected five vertices, approximately placing the return size halfway between the degree scales specified above. Performance is measured in edges returned/second.

All database operations were initiated through D4M in either MATLAB\textregistered{} or Julia 0.6 on the MIT SuperCloud \cite{reuther2013llsupercloud}, which consists of several 16-core Xeon-E5 machines with 64 GB of RAM. These operations were performed on a single node Accumulo instance running on the MIT SuperCloud dynamic database system~\cite{prout2015enabling}.

\section{Results}
\label{sec:results}

\begin{figure}[]
{\includegraphics[width=3.45in]{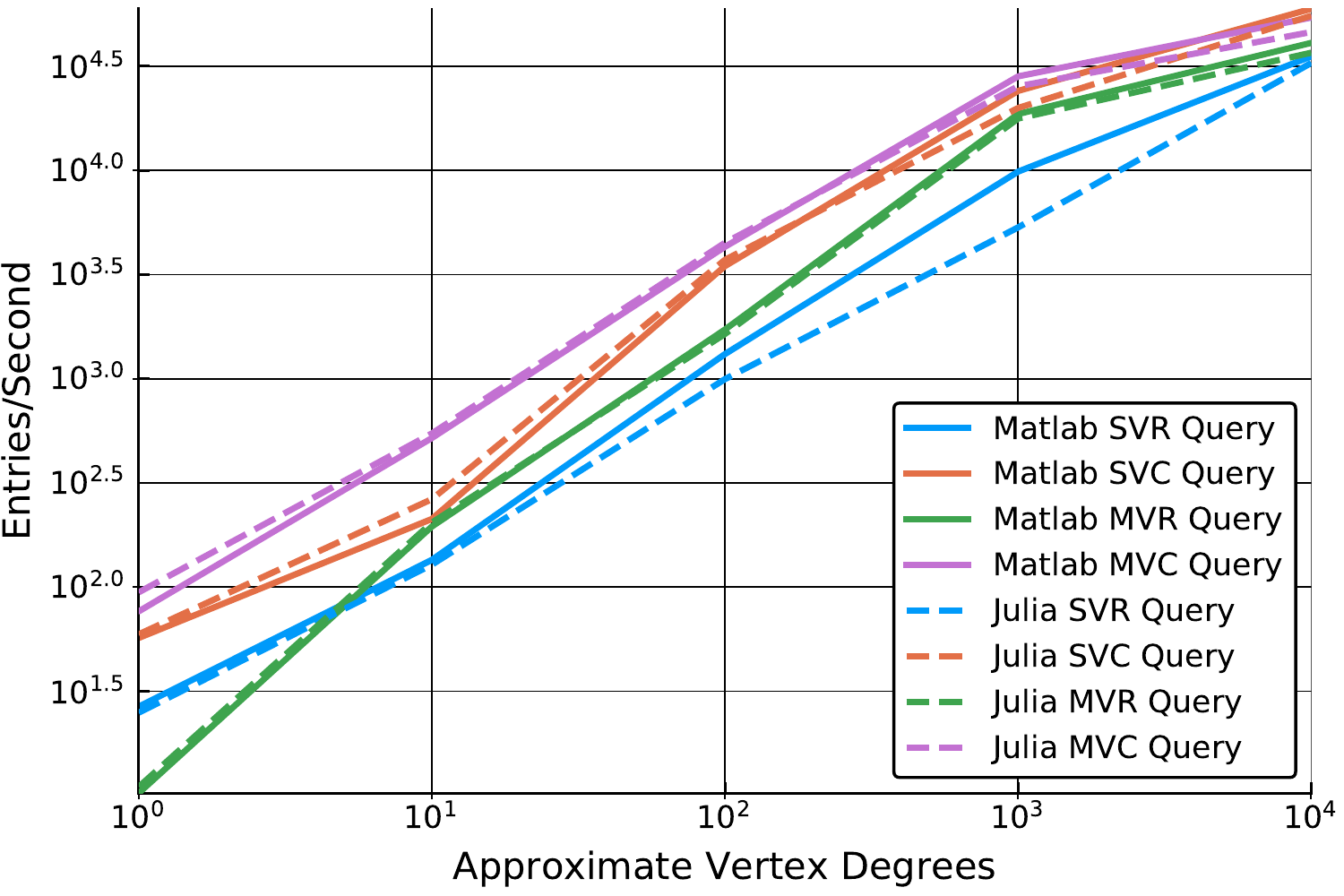}}
\centering
\caption{Query rates in entries returned/second against the degrees of the vertices queried for Matlab-D4M (solid lines) and D4M.jl (dashed lines). The queries run are single vertex row (SVR), single vertex column (SVC), multiple vertex row (MVR), and multiple vertex column (MVC).}
\label{fig:queryResults}
\end{figure}

Figures \ref{fig:ingestResults} and \ref{fig:queryResults} show performance results for ingest and query, respectively, comparing Matlab-D4M and D4M.jl. Figure \ref{fig:ingestResults} shows two views of ingest performance: the ingest rate against the number of ingest processes (left) and against the graph size (right).

The first plot in Figure \ref{fig:ingestResults} shows ingest rate scales with the number of ingest processes for four of the seven graph sizes. Performance for the remaining three were similar and were left out for readability. The Julia implementation ingested data at a faster rate than the Matlab-D4M in most cases, with the exception of the size 14 graph. The ingest rates increase fairly consistently with the increasing number of ingest processes, possibly dropping off somewhat at 16 processes on the larger graphs. The Matlab-D4M ingest processes see similar drop off, suggesting this may be Accumulo related.

In the second plot in Figure \ref{fig:ingestResults}, we see how the ingest rate scales with graph size. One thing to note is the lines corresponding to the D4M.jl ingest tend to be more flat and gently sloped than those of the Matlab-D4M ingest, with the exception of the 16 process ingest. The Matlab-D4M ingest starts at a lower rate than D4M.jl, peaks and surpasses the D4M.jl ingest rate for graphs of size ~13-15, and tapers off to a slower rate than D4M.jl at larger graph sizes. D4M.jl shows a similar increase and then decrease in ingest rate as graph size increases, but it is much less pronounced. The best ingest rates occurred for graphs of size 13 and 14. Both Julia and Matlab D4M ingest in batches with approximately 500,000 characters in each batch by default, which has previously been selected to give the best performance. At size 13 and 14, the entire graph fits into one batch, whereas size 15 and above must be inserted in two or more batches.

The improvement in ingest performance for D4M.jl may be attributed to better string array handling in Julia. On the D4M side, ingest mainly consists of extracting the triples from the Associative Array and splitting them into appropriate sized chunks before calling the Java function that ingests them. That the D4M.jl ingest rates are better than Matlab-D4M's at the larger graph sizes suggests that this may be the case.

Figure \ref{fig:queryResults} shows how the edges returned/second increases for increasing number of expected returned edges. The corresponding D4M.jl and Matlab-D4M lines in this plot follow each other  consistently, with D4M.jl faster in in some cases and Matlab-D4M in others. Both column queries returned more edges/second than the row queries, although in each case the multiple vertex queries returned more edges per second than their corresponding single vertex queries, as expected.

 \section{Conclusions and Future Work}
D4M is used for both its flexible data representation and manipulation and its ability to connect to a number of databases. In this work, we introduced D4M.jl database capabilities using D4M's custom Accumulo connector. We used the JavaCall Julia package to make calls to the Accumulo connector, which is written in Java. We provided simple to use wrapper functions that abstract away the Java function calls, allowing both flexibility and ease of use.

Overall results show that D4M.jl performs comparable to or better than Matlab-D4M. For the most part, D4M.jl seems to scale better with increasing the number of ingest processes, and scales to larger graph sizes much more gracefully than Matlab-D4M. In all cases, query rates for D4M.jl and Matlab-D4M were very close. Both the MATLAB\textregistered{}/GNU Octave and Julia implementations of D4M can be accessed through the D4M website download page \cite{d4mdownload}.

The next obvious step for this work is to provide the functionality to make Graphulo calls from Julia. Graphulo is a package that consists of server-side iterators for Accumulo that implement GraphBLAS kernels, which can be used to run graph algorithms on data stored in Accumulo. Like Matlab-D4M, we may be interested in adding interfaces to connectors for other types of databases, such as SQL and SciDB.

\section*{Acknowledgment}

The authors acknowledge the following individuals for their contributions: Chris Hill, Alan Edelman, Alex Chen, Dylan Hutchison, Bill Arcand, Bill Bergeron, David Bestor, Chansup Byun, Mike Houle, Matt Hubbell, Mike Jones, Anna Klein, Pete Michaleas, Julie Mullen, Andy Prout, Tony Rosa, and Chuck Yee. 



%
\IEEEtriggeratref{7}
\bibliography{D4M_HPEC_2018}

\begin{thebibliography}{10}
\providecommand{\url}[1]{#1}
\csname url@samestyle\endcsname
\providecommand{\newblock}{\relax}
\providecommand{\bibinfo}[2]{#2}
\providecommand{\BIBentrySTDinterwordspacing}{\spaceskip=0pt\relax}
\providecommand{\BIBentryALTinterwordstretchfactor}{4}
\providecommand{\BIBentryALTinterwordspacing}{\spaceskip=\fontdimen2\font plus
\BIBentryALTinterwordstretchfactor\fontdimen3\font minus
  \fontdimen4\font\relax}
\providecommand{\BIBforeignlanguage}[2]{{%
\expandafter\ifx\csname l@#1\endcsname\relax
\typeout{** WARNING: IEEEtran.bst: No hyphenation pattern has been}%
\typeout{** loaded for the language `#1'. Using the pattern for}%
\typeout{** the default language instead.}%
\else
\language=\csname l@#1\endcsname
\fi
#2}}
\providecommand{\BIBdecl}{\relax}
\BIBdecl

\bibitem{accumulo-webpage}
\BIBentryALTinterwordspacing
``Apache accumulo.'' [Online]. Available: \url{https://accumulo.apache.org/}
\BIBentrySTDinterwordspacing

\bibitem{accumulo-cloud}
S.~Zareian, M.~Fokaefs, H.~Khazaei, M.~Litoiu, and X.~Zhang, ``A big data
  framework for cloud monitoring,'' in \emph{Proceedings of the 2Nd
  International Workshop on BIG Data Software Engineering}, ser. BIGDSE
  '16.\hskip 1em plus 0.5em minus 0.4em\relax New York, NY, USA: ACM, 2016, pp.
  58--64.

\bibitem{accumulo-spatial}
K.~Lee, R.~K. Ganti, M.~Srivatsa, and L.~Liu, ``Efficient spatial query
  processing for big data,'' in \emph{Proceedings of the 22Nd ACM SIGSPATIAL
  International Conference on Advances in Geographic Information Systems}, ser.
  SIGSPATIAL '14.\hskip 1em plus 0.5em minus 0.4em\relax New York, NY, USA:
  ACM, 2014, pp. 469--472.

\bibitem{newsqlgraphulo2016}
D.~Hutchison, J.~Kepner, V.~Gadepally, and B.~Howe, ``From nosql accumulo to
  newsql graphulo: Design and utility of graph algorithms inside a bigtable
  database,'' in \emph{IEEE High Performance Extreme Computing (HPEC)}.\hskip
  1em plus 0.5em minus 0.4em\relax IEEE, 2016.

\bibitem{accumulo-bench}
R.~Sen, A.~Farris, and P.~Guerra, ``Benchmarking apache accumulo bigdata
  distributed table store using its continuous test suite,'' in
  \emph{Proceedings of the 2013 IEEE International Congress on Big Data}, ser.
  BIGDATACONGRESS '13, 2013, pp. 334--341.

\bibitem{ingest2014}
J.~Kepner, W.~Arcand, D.~Bestor, B.~Bergeron, C.~Byun, V.~Gadepally,
  M.~Hubbell, P.~Michaleas, J.~Mullen, A.~Prout, A.~Reuther, A.~Rosa, and
  C.~Yee, ``Achieving 100,000,000 database inserts per second using accumulo
  and d4m,'' in \emph{IEEE High Performance Extreme Computing (HPEC)}.\hskip
  1em plus 0.5em minus 0.4em\relax IEEE, 2014.

\bibitem{d4m2012}
J.~Kepner, W.~Arcand, W.~Bergeron, N.~Bliss, R.~Bond, C.~Byun, G.~Condon,
  K.~Gregson, M.~Hubbell, J.~Kurz, A.~McCabe, P.~Michaleas, A.~Prout,
  A.~Reuther, A.~Rosa, and C.~Yee, ``Dynamic distributed dimensional data model
  (d4m) database and computation system,'' in \emph{2012 IEEE International
  Conference on Acoustics, Speech and Signal Processing (ICASSP)}.\hskip 1em
  plus 0.5em minus 0.4em\relax IEEE, 2012, pp. 5349--5352.

\bibitem{julia2016}
A.~Chen, A.~Edelman, J.~Kepner, V.~Gadepally, and D.~Hutchison, ``Julia
  implementation of the dynamic distributed dimensional data model,'' in
  \emph{IEEE High Performance Extreme Computing (HPEC)}.\hskip 1em plus 0.5em
  minus 0.4em\relax IEEE, 2016.

\bibitem{accumuloEigensolver2015}
Y.~Huang, Y.~Yesha, and S.~Zhou., ``A database-based distributed computation
  architecture with accumulo and d4m: An application of eigensolver for large
  sparse matrix.'' in \emph{Big Data (Big Data), 2015 IEEE International
  Conference on. IEEE}, 2015.

\bibitem{cmd2015}
V.~Gadepally, B.~Hancock, B.~Kaiser, J.~Kepner, P.~Michaleas, M.~Varia, and
  A.~Yerukhimovich, ``Computing on masked data to improve the security of big
  data,'' in \emph{Technologies for Homeland Security (HST), 2015 IEEE
  International Symposium}.\hskip 1em plus 0.5em minus 0.4em\relax IEEE, 2015.

\bibitem{d4mschema2013}
J.~Kepner, C.~Anderson, W.~Arcand, D.~Bestor, B.~Bergeron, C.~Byun, M.~Hubbell,
  P.~Michaleas, J.~Mullen, D.~O'Gwynn, A.~Prout, A.~Reuther, A.~Rosa, and
  C.~Yee, ``D4m 2.0 schema: A general purpose high performance schema for the
  accumulo database,'' in \emph{IEEE High Performance Extreme Computing
  Conference (HPEC)}.\hskip 1em plus 0.5em minus 0.4em\relax IEEE, 2013.

\bibitem{juliapage}
\BIBentryALTinterwordspacing
``Julia.'' [Online]. Available: \url{https://julialang.org/}
\BIBentrySTDinterwordspacing

\bibitem{juliadatabases}
\BIBentryALTinterwordspacing
``Data base.'' [Online]. Available:
  \url{https://github.com/svaksha/Julia.jl/blob/master/DataBase.md}
\BIBentrySTDinterwordspacing

\bibitem{kepner2011}
J.~Kepner and J.~Gilbert, \emph{Graph algorithms in the language of linear
  algebr}.\hskip 1em plus 0.5em minus 0.4em\relax SIAM, 2011, vol.~22.

\bibitem{juliaFresh2014}
J.~Bezanson, A.~Edelman, S.~Karpinski, and V.~B. Shah, ``Julia: A fresh
  approach to numerical computing,'' \emph{arXiv preprint arXiv:1411.1607},
  2014.

\bibitem{juliaLang2012}
J.~Bezanson, S.~Karpinski, V.~B. Shah, and A.~Edelman, ``Julia: A fast dynamic
  language for technical computing,'' \emph{arXiv preprint arXiv:1209.5145},
  2012.

\bibitem{juliaHPC2016}
\BIBentryALTinterwordspacing
D.~Eadline. (2016) Dirt simple hpc: Making the case for julia. [Online].
  Available:
  \url{http://www.nextplatform.com/2016/01/26/dirt-simple-hpc-making-the-case-for-julia/}
\BIBentrySTDinterwordspacing

\bibitem{d4mjl2018}
\BIBentryALTinterwordspacing
``D4m.jl.'' [Online]. Available: \url{https://github.com/Accla/D4M.jl}
\BIBentrySTDinterwordspacing

\bibitem{javacall}
\BIBentryALTinterwordspacing
``Java call.'' [Online]. Available:
  \url{https://github.com/JuliaInterop/JavaCall.jl}
\BIBentrySTDinterwordspacing

\bibitem{pmatlab}
N.~T. Bliss, J.~Kepner, H.~Kim, and A.~Reuther, ``pmatlab: Parallel matlab
  library for signal processing applications,'' in \emph{2007 IEEE
  International Conference on Acoustics, Speech and Signal Processing - ICASSP
  '07}, vol.~4, April 2007, pp. IV--1189--IV--1192.

\bibitem{distarrays}
\BIBentryALTinterwordspacing
``Distributed arrays.'' [Online]. Available:
  \url{https://github.com/JuliaParallel/DistributedArrays.jl}
\BIBentrySTDinterwordspacing

\bibitem{d4mpowerlaw2015}
V.~Gadepally and J.~Kepner, ``Using a power law distribution to describe big
  data,'' in \emph{High Performance Extreme Computing Conference (HPEC)}.\hskip
  1em plus 0.5em minus 0.4em\relax IEEE, 2015.

\bibitem{reuther2013llsupercloud}
A.~Reuther, J.~Kepner, W.~Arcand, D.~Bestor, B.~Bergeron, C.~Byun, M.~Hubbell,
  P.~Michaleas, J.~Mullen, A.~Prout \emph{et~al.}, ``Llsupercloud: Sharing hpc
  systems for diverse rapid prototypingsupercloud: Sharing hpc systems for
  diverse rapid prototyping,'' in \emph{High Performance Extreme Computing
  Conference (HPEC), 2013 IEEE}.\hskip 1em plus 0.5em minus 0.4em\relax IEEE,
  2013, pp. 1--6.

\bibitem{prout2015enabling}
A.~Prout, J.~Kepner, P.~Michaleas, W.~Arcand, D.~Bestor, B.~Bergeron, C.~Byun,
  L.~Edwards, V.~Gadepally, M.~Hubbell \emph{et~al.}, ``Enabling on-demand
  database computing with mit supercloud database management system,'' in
  \emph{High Performance Extreme Computing Conference (HPEC), 2015 IEEE}.\hskip
  1em plus 0.5em minus 0.4em\relax IEEE, 2015, pp. 1--6.

\bibitem{d4mdownload}
\BIBentryALTinterwordspacing
``D4m downloads.'' [Online]. Available: \url{http://d4m.mit.edu/download}
\BIBentrySTDinterwordspacing

\end{thebibliography}
\bibliographystyle{IEEEtran}

\end{document}